%% file: main.tex
\documentclass[%
10pt,A4paper,twocolumn,aps,prx, reprint, superscriptaddress,longbibliography, nofootinbib,
 amsmath,amssymb,
]{revtex4-2}
\usepackage{lipsum,etoolbox}
\usepackage{mathtools}
\usepackage{graphicx}
\usepackage{dcolumn}
\usepackage{bm}
\usepackage{xcolor}
\usepackage{siunitx}
\usepackage{booktabs} 
\usepackage{array}
\usepackage{calrsfs}
\usepackage{multirow}
\usepackage{makecell}
\usepackage{enumitem}
\usepackage{placeins}
\usepackage{comment}
\usepackage{braket}
\usepackage{xspace}
\usepackage{tabularx}
\usepackage{array}
\usepackage{makecell} 
\usepackage{dpfloat}
\usepackage{refcount}
\usepackage{capt-of}
\usepackage{transparent}
\usepackage[export]{adjustbox}
\usepackage[shortcuts]{extdash}
\usepackage[colorlinks=true, linkcolor=blue, citecolor=black, urlcolor=blue, pdfauthor={Luca}, pdftitle={logical_state_teleportation_manuscript}]{hyperref}
\usepackage{cleveref}
\usepackage{upgreek}
\usepackage{circledsteps}
\pgfkeys{/csteps/inner xsep=2pt, /csteps/inner ysep=2pt}

\crefname{appendix}{app.}{apps.}
\Crefname{appendix}{App.}{Apps.}
\Crefname{figure}{Fig.}{Figs.}

\makeatletter
\def\@hangfrom@appendix#1#2#3{%
  #1%
  \@if@empty{#2}{%
    \MakeTextUppercase{#3}%
  }{%
    \MakeTextUppercase{#2}:\quad\MakeTextUppercase{#3}%
  }%
}
\def\@hangfroms@appendix#1#2{#1\MakeTextUppercase{#2}}
\makeatother

\DeclareMathAlphabet{\mathcal}{OMS}{cmsy}{m}{n}
\setlist[itemize]{leftmargin=*}

\input{memory_variables.tex}
\input{gate_variables.tex}
\input{protocol_variables.tex}

\input{step_variables.tex}
\input{leakage_rejection.tex}

\newcommand{\Lone}{\text{L1}} 
\newcommand{\Ltwo}{\text{L2}}
\newcommand{\Lfull}{\text{L}} 
\newcommand{\teleportation}{teleportation} 
 
\newcommand{\targetqubit}{target logical qubit}

\newcommand{\sourcequbit}{source logical qubit}

\newcommand{\XS}{\ensuremath{\hat{X}_\mathrm{\Lone}}}
\newcommand{\ZS}{\ensuremath{\hat{Z}_\mathrm{\Lone}}}
\newcommand{\XT}{\ensuremath{\hat{X}_\mathrm{\Ltwo}}}
\newcommand{\ZT}{\ensuremath{\hat{Z}_\mathrm{\Ltwo}}}
\newcommand{\XM}{\ensuremath{\hat{X}_\mathrm{\Lfull{}}}}
\newcommand{\ZM}{\ensuremath{\hat{Z}_\mathrm{\Lfull{}}}}
\newcommand{\XSXT}{\ensuremath{\hat{X}_\mathrm{\Lone}\hat{X}_\mathrm{\Ltwo}}}
\newcommand{\ZSZT}{$\hat{Z}_\mathrm{\Lone}\hat{Z}_\mathrm{\Ltwo}$}

\newlength{\imgw}
\newlength{\striph}
\newlength{\imgh}
\newlength{\sidemarg}
\newlength{\cutA}
\newlength{\margnat}
\newlength{\restA}
\newlength{\restB}
\newlength{\startB}
\makeatletter
\def\bibsection{%
  \par
  \begingroup
    \baselineskip26\p@
    \bib@device{\linewidth}{72\p@}%
  \endgroup
  \nobreak\@nobreaktrue
  \addvspace{19\p@}%
}
\makeatother

\begin{document}
\preprint{APS/123-QED}

\title{Teleportation of Logical Qubits using Lattice Surgery with Superconducting Circuits}

\newcommand{\affiliationeth}{\affiliation{Department of Physics, ETH Zurich, CH-8093 Zurich, Switzerland}}
\newcommand{\affiliationqc}{\affiliation{Quantum Center, ETH Zurich, CH-8093 Zurich, Switzerland}}
\newcommand{\affiliationpsi}{\affiliation{ETH Zurich - PSI Quantum Computing Hub, Paul Scherrer Institute, CH-5232 Villigen, Switzerland}}
\newcommand{\affiliationfzj}{\affiliation{Institute for Theoretical Nanoelectronics (PGI-2), Forschungszentrum Jülich, Jülich, Germany}}
\newcommand{\affiliationrwth}{\affiliation{Institute for Quantum Information, RWTH Aachen University, Aachen, Germany}}

\author{Luca~Hofele}\affiliationeth\affiliationqc
\author{Yongxin~Song}\affiliationeth\affiliationqc\affiliationpsi
\author{Áron~Márton}\affiliationfzj\affiliationrwth
\author{Ian~Hesner}\affiliationeth\affiliationqc\affiliationpsi
\author{Ilya~Besedin}\affiliationeth\affiliationqc\affiliationpsi
\author{Jakob~Ekert}\affiliationeth\affiliationqc\affiliationpsi
\author{Jonathan~Trebst}\affiliationeth\affiliationqc\affiliationpsi
\author{François~Swiadek}\affiliationeth\affiliationqc
\author{Felix~Henrich}\affiliationeth\affiliationqc\affiliationpsi
\author{Dante~Colao~Zanúz}\affiliationeth\affiliationqc\affiliationpsi
\author{Mohsen~Bahrami~Panah}\affiliationeth\affiliationqc\affiliationpsi
\author{Lukas~Bödeker}\affiliationfzj\affiliationrwth
\author{Nicholas~Gialouris}\affiliationeth\affiliationqc
\author{Chrysander~D.~Hagen}\affiliationeth\affiliationqc
\author{Marek~Pechal}\affiliationeth\affiliationqc\affiliationpsi
\author{Jean-Claude~Besse}\affiliationeth\affiliationqc\affiliationpsi
\author{Markus~Müller}\affiliationfzj\affiliationrwth
\author{Christoph~Hellings}\affiliationeth\affiliationqc
\author{Andreas~Wallraff}\affiliationeth\affiliationqc\affiliationpsi

\begin{abstract}
Demonstrating a quantum advantage when solving large-scale computational problems on a quantum computer requires the correction of errors that occur due to limitations in control accuracy and coherence. Using the concept of quantum error correction, logical qubits can be realized and fault-tolerantly manipulated with logical single- and two-qubit gates.
In this paper, we experimentally analyze fault-tolerant lattice-surgery, which is a key primitive that enables operations between logical qubits realized in the surface-code.
Using this primitive, we demonstrate the \teleportation{} of logical states between two distance-three logical qubits realized in a superconducting circuit composed of 41 physical qubits.
Throughout this process, we study the step-by-step transfer of logical information from a source to a target logical qubit by characterizing the time evolution of logical observables.
We find that repeated execution of stabilizer measurements, after merging two surface-code patches, restores fault tolerance and yields increased fidelity of the teleported state.
When optimizing the protocol in terms of gate count and sequence duration, we observe further improvements in fidelity.
Our work offers new insights into the optimization of fault-tolerant lattice-surgery operations, while also providing a comprehensive description of the physical device and gate sequences used. 
We expect the efficient implementation of increasingly complex logical circuits to enhance the performance of future fault-tolerant quantum computers.  
\end{abstract}

\date{\today}

\maketitle

\section{Introduction}
    Quantum error correction (QEC) protects logical qubits by redundantly encoding information across multiple physical qubits~\cite{Shor1995, Gottesman1997, Aharonov1997, Knill1998}. Provided physical error rates are below a code-specific threshold \cite{Aharonov1997}, scaling the number of physical qubits per logical qubit exponentially suppresses logical error rates, allowing logical qubits to achieve the fidelities required for large-scale, fault-tolerant quantum computation~\cite{Reiher2017, Preskill2018, Gidney2025b, Babbush2026}. The surface code~\cite{Kitaev2003, Dennis2002, Fowler2012} is a leading candidate toward that goal, offering planar nearest-neighbor connectivity, a high threshold of \SI{1}{\percent} against circuit-level noise~\cite{Raussendorf2007, Fowler2009}, and efficient, high-accuracy decoders~\cite{Higgott2025, Delfosse2021b}. These properties have motivated surface-code memory experiments that span initial proof-of-principle demonstrations~\cite{Krinner2022, Zhao2022d}, realizations of logical error suppression~\cite{Acharya2023, Acharya2025, He2025c}, incorporation of adaptive control techniques~\cite{Sivak2026}, and mitigation of defective components~\cite{Drouet2026}.\par
    
    Extending quantum memory toward universal quantum computation requires implementing logical operations~\cite{Gottesman2010}. 
    Lattice surgery is a technique for realizing such operations using joint logical Pauli measurements implemented through code deformation~\cite{Horsman2012, Bombin2009}. These measurements are performed by merging and splitting patches of logical qubits via adjacent merging regions \cite{Bodeker2026} using the same lattice structure that encodes the logical qubits. 
    Lattice surgery serves as a building block for both logical Clifford operations and magic state distillation~\cite{Fowler2018, Gidney2019, Litinski2019}, thereby enabling a universal set of logic gates. 
    
    First experiments of lattice surgery have been demonstrated with trapped ions \cite{Erhard2021, Ryan-Anderson2024, Steiner2026}, Rydberg atoms \cite{Bluvstein2025a}, and superconducting circuits implementing color codes \cite{Lacroix2025b}, repetition codes \cite{Besedin2026}, surface codes \cite{Wang2026o, Lin2026c, Lange2026, Harper2026a}, dynamic compass codes \cite{McLauchlan2026}, and Bacon-Shor codes \cite{Hetenyi2024}.
    While these experiments focused primarily on the logical fidelity of the output state, here, we systematically characterize the time evolution of logical expectation values during lattice surgery operations between two distance-three logical qubits. We experimentally demonstrate fault-tolerant \teleportation{} from a \sourcequbit{} to a \targetqubit{}.
    By evaluating logical observables at different steps of the protocol, we experimentally quantify how repeated stabilizer measurements following a lattice merge operation reduce logical errors, underscoring the importance of fault tolerance for high-fidelity logical operations. 
    Retaining only the stabilizer measurements required to preserve fault tolerance in \teleportation, we execute two optimized circuits with fewer stabilizer measurements \cite{Zhou2025f, Hangleiter2025}.
    \begin{figure*}[t!]
		\includegraphics[width=\linewidth]{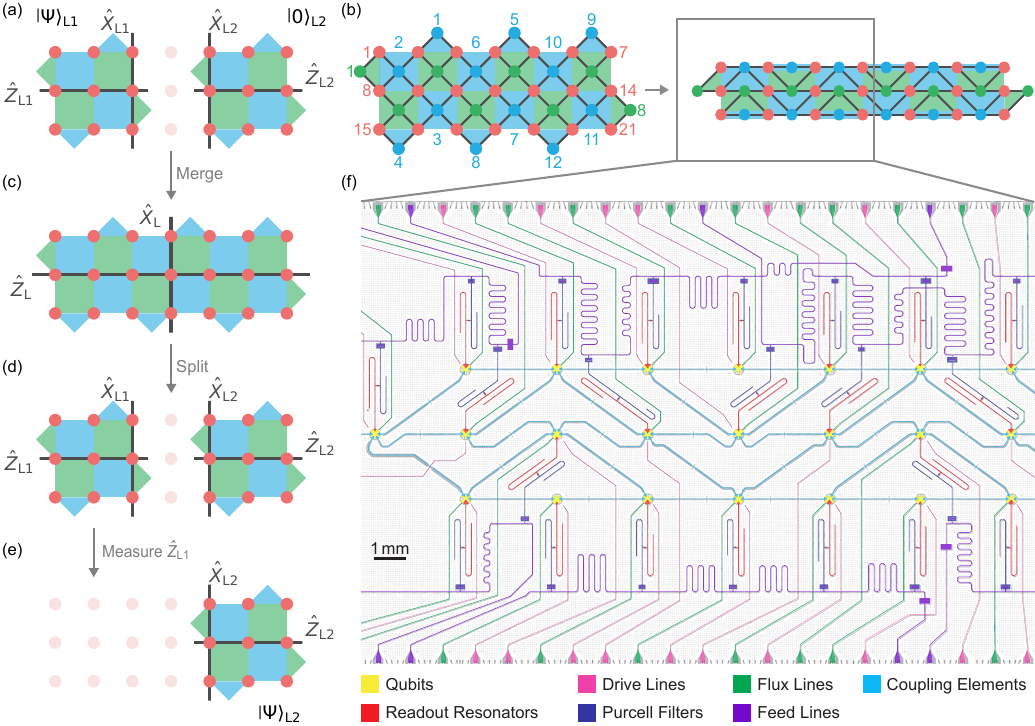}
		\caption{\label{fig:fig1_quantum_device}
			Lattice surgery operations to implement logical teleportation on a superconducting qubit device. In the schematic illustrations, data qubits are shown as red circles, $Z$-type ($X$-type) stabilizers are indicated by green (blue) plaquettes, and logical operators by black lines.
            (a) Two logical qubits (\Lone{} and \Ltwo{}) with logical operators \ZS{}, \XS{}, \ZT{}, and \XT{}, initialized in $\ket{\Psi}_\Lone{}$ and $\ket{0}_\Ltwo{}$ as in the logical teleportation protocol implemented in this paper.
			(b) Mapping of a regular three-by-seven surface code lattice to the qubit arrangement on the quantum device. 
            In the right lattice, weight-two $X$ stabilizers are drawn as thick blue lines.
            Coupling between qubits are indicated with black lines. Indices are from D1 to D21 for data qubits (red), from X1 to X12 for $X$-type auxiliary qubits (blue), and from Z1 to Z8 for $Z$-type auxiliary qubits (green). 
            (c) Three-by-seven surface code \Lfull{} with logical operators \ZM{} and \XM{} resulting from a merge operation (gray arrow) applied to \Lone{} and \Ltwo{}.
            (d) \Lone{} and \Ltwo{} after applying a split operation (gray arrow) to \Lfull{}. 
            (e) Logical qubit \Ltwo{} in a final state $\ket{\Psi}_\Ltwo{}$ after an individual logical measurement of \Lone{} (gray arrow) and conditional Pauli operations (not shown).
            (f) False-color optical micrograph of the section of the quantum device indicated by the gray rectangle in (b).
		}
	\end{figure*}
    We compare the output state fidelity of teleported Pauli $X$ and $Z$ eigenstates for a fully modular and two optimized circuits, demonstrating that reducing stabilizer measurements in quantum error correction protocols improves the logical performance.
    
	\section{Modular Logical qubits}\label{sec:concepts}
    Using nine data qubits (red circles) each, we encode two distance-three logical qubits (\Lone{} and \Ltwo) in the surface code. We use eight auxiliary qubits per logical qubit to measure the stabilizers of the code (green and blue plaquettes) to detect and correct errors, see \Cref{fig:fig1_quantum_device}(a).
    We define the logical operators of \Lone{} and \Ltwo{} as ${\hat{Z}_\Lone=\hat{Z}_\mathrm{D8}\hat{Z}_\mathrm{D9}\hat{Z}_\mathrm{D10}}$, ${\hat{X}_\Lone=\hat{X}_\mathrm{D3}\hat{X}_\mathrm{D10}\hat{X}_\mathrm{D17}}$, ${\hat{Z}_\Ltwo=\hat{Z}_\mathrm{D12}\hat{Z}_\mathrm{D13}\hat{Z}_\mathrm{D14}}$ and ${\hat{X}_\Ltwo=\hat{X}_\mathrm{D5}\hat{X}_\mathrm{D12}\hat{X}_\mathrm{D19}}$, as indicated by the black lines in \Cref{fig:fig1_quantum_device}(a). The numbering of the data qubits is shown in panel~(b). 
    At the interface between \Lone{} and \Ltwo, four auxiliary qubits ($\mathrm{X5, X6, X7, X8}$), labeled in blue in \Cref{fig:fig1_quantum_device}(b), and three data qubits ($\mathrm{D4, D11, D18}$) are used to merge the logical qubits into a single three-by-seven surface code \Lfull{}, see \Cref{fig:fig1_quantum_device}(c). After splitting \Lfull{} back into \Lone{} and \Ltwo{}, see \Cref{fig:fig1_quantum_device}(d), the modularity of the logical qubits allows us to measure \Lone{} individually, see \Cref{fig:fig1_quantum_device}(e), which enables logical teleportation as discussed in \Cref{sec:teleportation}. \par
    
    We have designed and fabricated a 41-qubit device in a planar architecture using an arrangement with three rows of qubits, alternating between data and auxiliary qubits within each row, see \Cref{fig:fig1_quantum_device}(b).
    Physical qubits are realized as frequency-tunable transmon qubits~\cite{Koch2007} (yellow) with static coplanar waveguide couplers (cyan).
    Each qubit has a dedicated drive (pink) and flux line (green) as well as a dedicated readout resonator (red) with a Purcell filter (dark blue) coupled to a common feed line (purple), see the false-color optical micrograph in \Cref{fig:fig1_quantum_device}(f). The common feed line is used for frequency-multiplexed, simultaneous readout of up to six qubits~\cite{Blais2004, Chen2012f, Jeffrey2014, Heinsoo2018}. A micrograph of the full device is provided in \Cref{app:micrograph}. \par
    
    To assess the performance of the device, we benchmark all physical operations relevant to lattice surgery. We realize single-qubit gates, with a mean gate error of \SQG{} characterized by randomized benchmarking (RB), see \Cref{app:gate_sequence}. We perform two-qubit operations based on flux-activated $\mathrm{C}Z$ gates~\cite{Strauch2003, DiCarlo2010, Negirneac2021}, with a mean gate error of \TQG{} determined by interleaved RB. Using dispersive readout, we achieve mean assignment errors of \ROtwo{} for the two and \ROthree{} for three lowest eigenstates of the transmon qubits~\cite{Blais2004, Chen2012f, Jeffrey2014, Heinsoo2018}. We combine these operations to perform eight weight-two and twelve weight-four stabilizer measurements of $X$ (blue plaquettes) and $Z$ (green plaquettes) type, as shown in \Cref{fig:fig1_quantum_device}(c), respectively. 

	\begin{figure}[bp]
		\includegraphics[width=\linewidth]{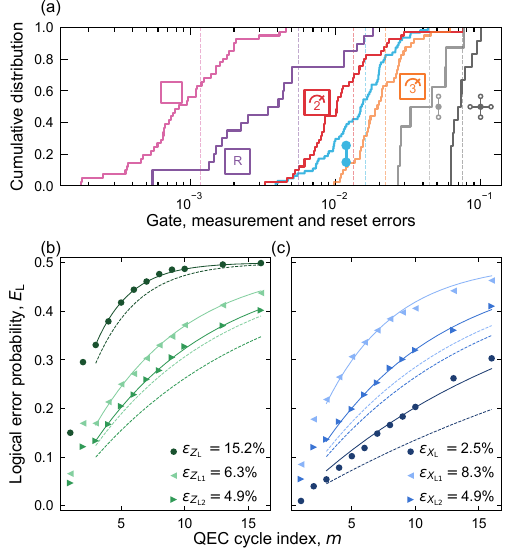}
		\caption{\label{fig:fig2_mean_syndromes}
			Experimental device characterization. 
            (a) Cumulative distributions of experimentally extracted errors in single-qubit gates (pink), two-qubit gates (cyan), reset (purple), two-state (red) and three-state (orange) readout, as well as weight-two (light-gray) and weight-four (dark-gray) stabilizers. Vertical dashed lines indicate the average error for each type of operation. 
            (b,c) Individual characterization of logical qubits \Lone{}, \Ltwo{}, and \Lfull{}. Logical error probabilities $E_\mathrm{L}$ (markers) as a function of the number of QEC cycles $m$ for (b) the logical operators \ZS, \ZT, \ZM{} and (c) \XS, \XT, \XM{}, see text for details. Solid lines are exponential fits to the measured data, starting at $m=3$ to reduce boundary effects~\cite{Acharya2023}. Dashed lines are fits to data generated in stabilizer simulations, see \Cref{app:appendix_sims}.
		}
	\end{figure}
    
    To characterize each stabilizer measurement individually, we sequentially prepare the data qubits $\mathrm{D}_j$ across all possible product states in the basis of the stabilizer, obtaining mean errors of \weightTwo{} for weight-two and \weightFour{} for weight-four stabilizers. Finally, we unconditionally reset the auxiliary qubits using a parametric flux modulation~\cite{Zhou2021, Lacroix2025} and achieve a mean residual population of \reseterror{} when preparing qubits in the first excited state, see \Cref{fig:fig2_mean_syndromes}(a). \par
    
    As a benchmark of building blocks on the logical level, we execute quantum memory experiments~\cite{Ryan-Anderson2021, Krinner2022, Zhao2022d} on the two logical qubits \Lone{} and \Ltwo{} individually. 
    We discard runs where leakage is detected in any readout performed during the experiment~\cite{Krinner2022}, retaining \SI{95.5}{\percent} of the runs per cycle for \Lone{} and \SI{94.5}{\percent} for \Ltwo{} (see \Cref{app:leakage}).

    At the beginning of each experimental run, we prepare the physical qubits by first reading out their initial states, which determines the initial stabilizer values for decoding, and then applying single-qubit rotations to initialize the logical state.
    Using minimum-weight perfect matching decoding with weights extracted from measured syndrome statistics~\cite{Spitz2018, Remm2026}, we extract the logical error probabilities as a function of the number of QEC cycles $m$, averaged over both eigenstates of each basis, see \Cref{app:gate_sequence} for details.
    From exponential fits, we extract logical error rates $\epsilon_Z$ and $\epsilon_X$ of \leftZ{} and \leftX{} for \Lone{}, and \rightZ{} and \rightX{} for \Ltwo{}, see \Cref{fig:fig2_mean_syndromes}(b) and (c).
    We attribute the observed performance difference between the two logical qubits to strongly coupled two-level-system defects~\cite{ColaoZanuz2025} for which occurrences vary across the device.

\section{Logical teleportation}\label{sec:teleportation}
\begin{figure}[!t]
		\centering
		\includegraphics[width=\linewidth]{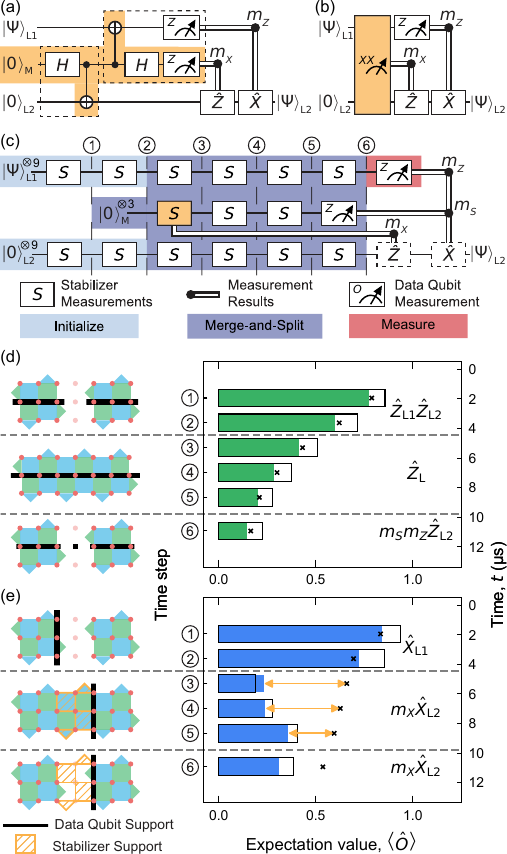}
		\caption{\label{fig:fig3_logical_state_teleportation}
            Time evolution of logical observables during \teleportation.
            (a) Original \teleportation{} protocol \cite{Bennett1993} using Hadamard (H) and CNOT gates for Bell state preparation and measurement (dashed boxes), and Pauli operations conditioned on measurement outcomes $m_Z$ and $m_X$ (double-lines). 
            (b) Equivalent circuit representing the operations highlighted in orange in (a) as an effective $XX$ parity measurement.
            (c) Lattice surgery implementation of (b). Top and bottom rows denote source and target qubits; the middle row represents their interface. Vertical dashed lines mark time steps. Combined $X$- and $Z$-type stabilizer measurements are indicated by boxes labeled $S$. 
            The stabilizer measurements indicated with an orange box implement the $XX$-parity measurement. Dashed boxes show Pauli frame updates, which we perform in post-processing. 
            (d, e) Evolution of logical observables across the protocol in (c). Pictograms show surface code lattices and their active stabilizers at each step in (c). Black lines and orange patches indicate support of logical observables (see legend). Colored bars show error-corrected expectation values; wireframes show simulation results. Black crosses indicate calculations based on errors only on the data qubit support, see \Cref{app:spacelike_model}. Orange arrows indicate differences between these calculations and the measured data.
        }
	\end{figure}

    We experimentally characterize the teleportation of a quantum state from logical qubit \Lone{} to logical qubit \Ltwo{} using lattice surgery. The original teleportation protocol \cite{Bennett1993}, independently of whether it acts on physical \cite{Steffen2013} or logical qubits, transfers an unknown quantum state $\ket{\Psi}$ from a source qubit to a target qubit, in this case from \Lone{} to \Ltwo{}. In the process, an additional auxiliary qubit~M is prepared in a maximally-entangled state with the target qubit \Ltwo{}, forming a Bell pair. Then qubits \Lone{} and M are measured in the Bell basis, and the measurement outcomes $m_Z$ and $m_X$ are used to complete the teleportation protocol by conditionally applying the operators \ZT{} and \XT{} to the target qubit \Ltwo, see \Cref{fig:fig3_logical_state_teleportation}(a). The circuit in the orange
    box in (a) constitutes a joint \XSXT{} measurement of \Lone{} and \Ltwo{}, see panel (b), which, when followed by the measurement of \Lone{} and the same feedforward steps as in (a), realizes teleportation.

    The circuit in \Cref{fig:fig3_logical_state_teleportation}(b) can be implemented directly using lattice surgery by extracting the joint logical operator \XSXT{} from stabilizer measurements during a merge operation and measuring the operator \ZS{} of the source logical qubit after a split operation, see \Cref{fig:fig1_quantum_device}(a).
        
    Below, we present a fully modular implementation of this protocol, operating on logical qubits defined on separate surface-code patches and constructed from fault-tolerant initialization, merge-and-split and measurement modules (\Cref{app:modular_FT}).
    Using this modular approach, we successfully teleport logical states in the $Z$ basis with probability \Zsuccess{} and in the $X$ basis with probability \Xsuccess{}. In \Cref{sec:optimization}, we present experimental results demonstrating fidelity improvements to the teleportation protocol using optimized circuits~\cite{Bodeker2026}.
    
    In the modular implementation of the teleportation protocol, we first initialize \Lone{} and \Ltwo{} by performing two rounds of stabilizer measurements, see light blue boxes in \Cref{fig:fig3_logical_state_teleportation}(c). We then merge \Lone{} and \Ltwo{} into \Lfull{} by 
    performing measurements of the stabilizers $\hat{S}_\mathrm{X5}, \hat{S}_\mathrm{X6}, \hat{S}_\mathrm{X7}, \hat{S}_\mathrm{X8}$ at the interface between the two logical qubits, yielding the measurement outcome $m_X$ of the observable $\hat{X}_\Lone{}\hat{X}_\Ltwo{}=\hat{S}_\mathrm{X5}\hat{S}_\mathrm{X6}\hat{S}_\mathrm{X7}\hat{S}_\mathrm{X8}$. 
    After two further rounds of stabilizer measurements required for fault-tolerance, see \Cref{app:modular_FT}, we split \Lfull{} back into \Lone{} and \Ltwo{}. We do this by reading out the data qubits D4, D11, and D18 at the interface and discontinuing the measurements of the stabilizers $\hat{S}_\mathrm{X5}, \hat{S}_\mathrm{X6}, \hat{S}_\mathrm{X7}, \hat{S}_\mathrm{X8}$, see the purple box in \Cref{fig:fig3_logical_state_teleportation}(c).
    By reading out the data qubits of the source logical qubit \Lone{}, we then obtain the measurement outcome~$m_Z$ of the logical operator \ZS{}. 
    The teleportation protocol is concluded by applying Pauli frame updates on the target logical qubit \Ltwo{} conditioned on $m_X, m_Z$ and on the result $m_S$ of reading out data qubit D11 in the split operation, see \Cref{fig:fig3_logical_state_teleportation}(c). \par

\section{Time evolution of logical observables}
    To characterize errors that occur during the logical teleportation protocol, we measure the expectation values of the observables indicated in Figs.~\ref{fig:fig3_logical_state_teleportation}(d) and (e) at each of the steps of the gate sequence in Fig.~\ref{fig:fig3_logical_state_teleportation}(c). The observables are chosen such that all faults that result in a logical error on the final teleported state also alter the intermediate observable. 
    During initialization, the logical $Z$ observable is given by the combined operator \ZSZT{} of both logical qubits while the logical $X$ observable is ${\hat{X}_\Lone}$, see the black lines in the pictograms of \Cref{fig:fig3_logical_state_teleportation}(d,e). The merge step transforms the logical $Z$ observable to the logical operator of the three-by-seven surface code, ${\hat{Z}_\mathrm{L}=\hat{Z}_\mathrm{D8}...\hat{Z}_\mathrm{D14}}$.
    From this point on, the logical $X$ observable is equal to ${\hat{X}_\Ltwo}$ adjusted by the Pauli frame update determined from the measurement result $m_X$ of the stabilizers at the interface of the logical qubits, see the orange plaquettes in the pictograms of \Cref{fig:fig3_logical_state_teleportation}(e).
    After the split operation, the logical $Z$ observable is equal to the logical operator \ZT{} of the \targetqubit{} adjusted by the Pauli frame update conditioned on the measurement results $m_S$ and $m_Z$. 

    We examine the time evolution of the logical $Z$ observable when teleporting the logical state $\ket{0}_\Lone$ (data for $\ket{1}_\Lone$ in \Cref{app:time_evolution_appendix}).
    We measure the product of the physical Pauli operators defining the logical observable at each time step and determine its expectation value. After leakage rejection and decoding with minimum-weight perfect matching, we find {$\langle\hat{Z}_\Lone\hat{Z}_\Ltwo\rangle= \ZstepOne$} (with ideal value 1) in the first step of the initialization \Circled{1} and {$\langle m_S m_Z\hat{Z}_\Ltwo\rangle= \ZstepSix$}, corresponding to a success probability of \Zsuccess{}, at the end of the teleportation protocol \Circled{6}, see green bars in \Cref{fig:fig3_logical_state_teleportation}(d). 
        
    The reduction of logical fidelity in consecutive steps of the protocol is consistent with the logical error rates extracted from the memory experiments on the two logical qubits and the merged three-by-seven surface code. Expectation values calculated from these rates, see \Cref{app:spacelike_model}, are shown as black crosses in \Cref{fig:fig3_logical_state_teleportation}(d). We also find reasonable agreement between our measured data and stabilizer simulations (wireframes in \Cref{fig:fig3_logical_state_teleportation}(d)) based on a Pauli noise model (see \Cref{app:appendix_sims}) incorporating the inhomogeneous errors for gates, readout and reset indicated in \Cref{fig:fig2_mean_syndromes}(a). 
    To avoid overinterpreting the predictive power of the simplified depolarizing noise model, we include neither biases of the Pauli channels nor errors from undetected leakage, residual $ZZ$ coupling or measurement-induced dephasing. These choices are the likely origin of the remaining differences between the data and the model.

    Unlike the logical $Z$ observable, the logical $X$ observable depends on stabilizer measurement outcomes, see the pictograms of \Cref{fig:fig3_logical_state_teleportation}(e), leading to a qualitatively different time evolution.
    We study this evolution by initializing the \sourcequbit{} in the $\ket{+}_\Lone$ state (data for $\ket{-}_\Lone$ in \Cref{app:time_evolution_appendix}).
    During initialization in steps~\Circled{1} and \Circled{2}, we obtain expectation values of {$\langle\hat{X}_\Lone\rangle=\XstepOne$} and \XstepTwo{}, see the blue bars. These values are in good agreement with errors on the data qubit support, see black crosses in \Cref{fig:fig3_logical_state_teleportation}(e), which are calculated in \Cref{app:spacelike_model} based on logical error rates extracted from the memory experiments in \Cref{fig:fig2_mean_syndromes}(c).
    The drop of the expectation value to {$\langle m_X\hat{X}_\Ltwo\rangle=\XstepThree$} in step~\Circled{3} exceeds what can be explained by errors on the data qubit support alone. We identify the additional errors (orange arrows) as errors occurring on the stabilizer support, i.e., errors on stabilizers at the interface between the two logical qubits (orange plaquettes).
    During the execution of steps~\Circled{4} and \Circled{5}, the errors on the stabilizer support reduce while the expectation value recovers to $\langle m_X\hat{X}_\Ltwo\rangle=\XstepFive$ and the protocol regains fault tolerance, see discussion in \Cref{app:modular_FT}.
    Stabilizer simulations (black wireframes) qualitatively confirm this distinctive pattern in the evolution of the expectation value.
    At the end of the teleportation protocol \Circled{6}, the expectation value is {$\langle m_X\hat{X}_\Ltwo\rangle=\XstepSix$}, equivalent to a success probability of \Xsuccess{} of the logical teleportation protocol in the $X$ basis.
    \par

    Our study of the evolution of logical observables over the course of the teleportation protocol reveals that the logical $X$ observable is, from the merge step onward, affected significantly by errors on its stabilizer support. 
    These errors are suppressed by repeated rounds of stabilizer measurements that restore fault tolerance, which highlights the performance gains of fault-tolerant circuits over non-fault-tolerant ones. 
    Despite the additional errors on the stabilizer support in the $X$ basis, we find a higher logical fidelity for teleporting $\ket{+}_\Lone$ and $\ket{-}_\Lone$ than for teleporting $\ket0_\Lone$ and $\ket1_\Lone$. 
    This asymmetry is due to the higher logical error per step on the data qubit support in the $Z$ basis, which is revealed by our analysis of the time evolution and discussed further in \Cref{app:spacelike_model}. 
    
\section{Optimization of Lattice Surgery Protocols}\label{sec:optimization}
    Constructing circuits from fault-tolerant elementary modules such as initialization, merge-and-split, and logical measurement ensures fault tolerance of the overall circuit.
    However, relaxing the requirement of individual fault tolerance of every elementary module can improve logical fidelity by reducing gate count and circuit duration \cite{Fowler2018, Gidney2019, Litinski2019, Hangleiter2025, Bodeker2026}.
    To optimize the logical teleportation circuit described in the previous section, we combine three elementary modules into a single larger one, as explained below, allowing for a reduction of the number of stabilizer measurements within the composite module while, importantly, retaining its fault tolerance.\par
    
\begin{figure}[t]
    		\centering
    		\includegraphics[width=\linewidth]{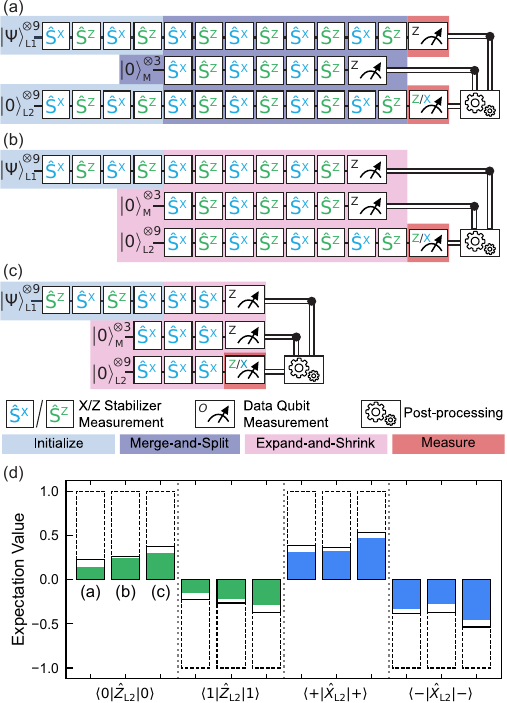}
    		\caption{\label{fig:fig4}
    			Fully modular vs.\ optimized logical \teleportation{} circuits.
    			(a)--(c) Sequences of stabilizer measurements and data qubit readout in three distinct circuits implementing logical \teleportation{} of $\ket{1}_\Lone{}, \ket{0}_\Lone{}, \ket{{+}}_\Lone{}, \ket{{-}}_\Lone{}$. Background colors in the circuits indicate QEC modules (see legend). 
                (d) Measured (bars), simulated (solid wireframes), and ideal (dashed wireframes) error-corrected Pauli expectation values of the indicated teleported states. Letters indicate the circuit used for each bar and are repeated across groups.
    		} 
    	\end{figure}

    In the modular \teleportation{} protocol, the initialization of the target logical qubit \Ltwo{} (bottom blue box in \Cref{fig:fig4}(a)), the merge-and-split module (purple box), and the logical measurement of the source qubit \Lone{} (top red box), effectively expand the source logical qubit into a three-by-seven surface code before shrinking it again to the lattice constituting the target logical qubit.
    Combining these three modules allows us to omit stabilizer measurements in the resulting expand-and-shrink module, indicated by the pink box in \Cref{fig:fig4}(b), while maintaining its fault tolerance (\Cref{app:modular_FT}).

    For this optimized circuit, we observe absolute expectation values of {$|\langle m_S m_Z\hat{Z}_\Ltwo\rangle|= \fidMoveZ$} and {$|\langle m_X\hat{X}_\Ltwo\rangle|=\fidMoveX$} averaged over $\ket{0}_\Lone{}, \ket{1}_\Lone{}$ and $\ket{+}_\Lone{}, \ket{-}_\Lone{}$, respectively.
    The improvement of \ModularToMoveZ{} in the $Z$ basis compared to the circuit in \Cref{fig:fig4}(a), see the green bars in \Cref{fig:fig4}(d), stems from depleting the modules for target-qubit initialization and source-qubit measurement.
    Since both modules are susceptible only to errors in the $Z$ basis (\Cref{app:basis_dependence}) the performance in the $X$ basis remains nearly unchanged, see the blue bars in \Cref{fig:fig4}(d).

    To further optimize the circuit, we note that $\hat{S}^{Z}$ stabilizer measurements are not required for fault-tolerance of a distance-three expand-and-shrink module (\Cref{app:modular_FT}).
    Omitting these stabilizer measurements reduces the number of errors introduced during syndrome extraction but also allows for errors in the $Z$ basis to accumulate over a longer period~\cite{Marton2025a}. 
    In addition, we allow a temporal overlap of the expand-and-shrink module with the source-qubit initialization and the target-qubit measurement, and we omit $\hat{S}^{X}$ stabilizer measurements that are otherwise scheduled at these points of the circuit but are not needed to achieve fault tolerance (\Cref{app:modular_FT}).

    In the resulting circuit in \Cref{fig:fig4}(c),
    teleporting $\ket{0}_\Lone{}, \ket{1}_\Lone{}$ and $\ket{+}_\Lone{}, \ket{-}_\Lone{}$ 
    yields average expectation values of {$|\langle m_S m_Z\hat{Z}_\Ltwo\rangle|= \fidDepletedZ$} and {$|\langle m_X\hat{X}_\Ltwo\rangle|=\fidDepletedX$}, representing significant improvements to the values of \fidTeleportZ{} and \fidTeleportX{} achieved with the circuit in \Cref{fig:fig4}(a).
    While the optimal degree of circuit depletion generally depends on the experimental error rates of different physical processes \cite{Fowler2018, Gidney2019, Litinski2019, Hangleiter2025, Marton2025a, Bodeker2026}, the depletion performed in \Cref{fig:fig4}(c) significantly improves the performance in both the $Z$ and $X$ basis, realizing logical teleportation with success probabilities of \successDepletedZ{} and \successDepletedX{}, respectively.
    
    Our results underscore a design trade-off for fault-tolerant protocols, balancing the versatility of elementary modules against the fidelity gains achieved through optimization of composite modules.
    
	\section{Conclusion and discussion} 
    We have demonstrated fault-tolerant logical teleportations of $Z$ and $X$ eigenstates in the surface code, with success probabilities up to \successDepletedZ{} and \successDepletedX{}, respectively.
    By analyzing the expectation values of logical observables at different steps of a lattice surgery protocol, we have experimentally observed a reduction in logical fidelity expected from a temporary breakdown of fault tolerance after a merge operation, followed by its restoration through repeated rounds of stabilizer measurements. 
    In future work, this analysis could be applied to the teleportation of arbitrary quantum states and extended to full tomography using a fault-tolerant $Y$-basis measurement~\cite{Gidney2024, Lin2026c} to gain further insights into error sources in lattice surgery protocols.
    Moreover, studying the evolution of logical observables in larger circuits, such as logical CNOT gates, and in non-Clifford routines, including magic-state cultivation~\cite{Gidney2024a, Rosenfeld2026}, provides a powerful diagnostic for optimizing complex QEC protocols. 
    Guided by the resulting insights, circuit optimization can be further advanced by refining depletion strategies in modular QEC circuits. Scaling these methods to higher code distances and error rates below the fault tolerance threshold can help to achieve the logical fidelities required for practically useful error-corrected quantum computation.

\section*{Acknowledgments}
    \vspace{-2ex}
    We thank E.~Baril and A.~Blais for their simulations of components of the device, K.~Bensink for his contributions to the control software, M.~Kerschbaum and R.~Keller for their contributions to the experimental setup, and A.~Remm for contributions during early stages of the project. We are grateful for valuable discussions with L.~Colmenarez. 
    This research was sponsored by the Army Research Office and was accomplished under Cooperative Agreement Number W911NF-23-2-0212, by the Swiss State Secretariat for Education, Research and Innovation (SERI) under contract number UeM019-11, and by ETH Zurich.
    The views and conclusions contained in this document are those of the authors and should not be interpreted as representing the official policies, either expressed or implied, of the Army Research Office or the U.S. Government. The U.S. Government is authorized to reproduce and distribute reprints for Government purposes notwithstanding any copyright notation herein. We acknowledge computing time provided at the NHR Center NHR4CES at RWTH Aachen University (Project No. p0020074) (A.M. and M.M.). This is funded by the Federal Ministry of Education and Research and the stategovernments participating on the basis of the resolutions of the GWK for national high-performance computing at universities. M.M.  furthermore acknowledges support by the Deutsche Forschungsgemeinschaft (DFG, German Research Foundation) under Germany’s Excellence Strategy Cluster of Excellence Matter and Light for Quantum Computing (ML4Q) EXC 2004/1 390534769.

\appendix
\crefalias{section}{appendix}

\section{Optical micrographs of the quantum device}\label{app:micrograph}

    False-color optical micrographs of the left and right parts of the device are shown in \Cref{fig:s43_wide_1} and \Cref{fig:s43_wide_2}, respectively.
    
    \settowidth{\imgw}{\includegraphics{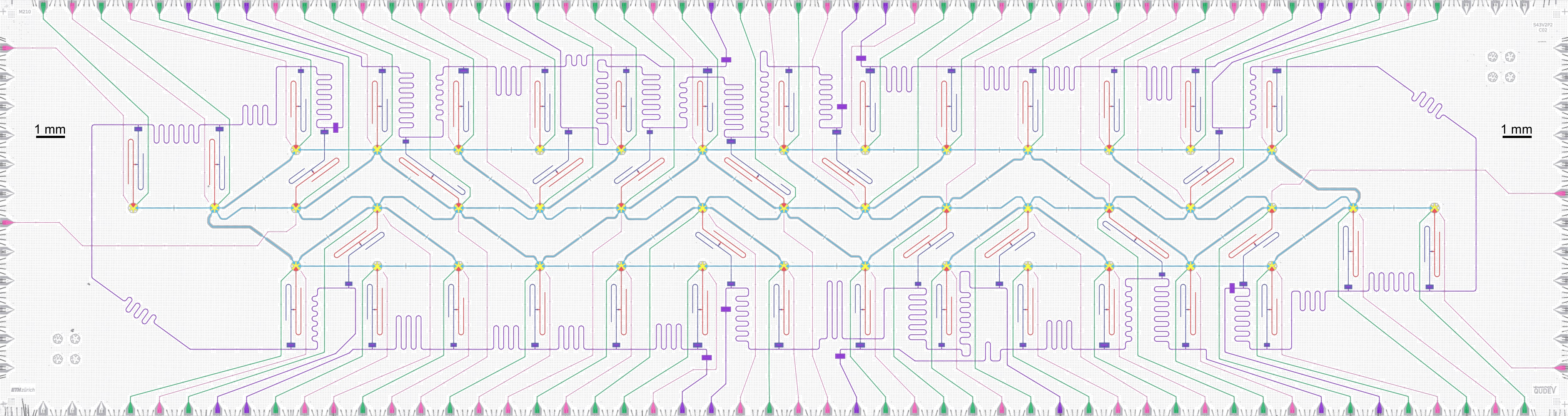}}
    \settoheight{\imgh}{\includegraphics{figures/device.pdf}}
    \setlength{\striph}{0.53\textwidth}                                  
    \setlength{\sidemarg}{\dimexpr(\paperwidth-\textwidth)/2\relax}      
    \setlength{\cutA}{\dimexpr\textwidth*\imgh/\striph\relax}            
    \setlength{\margnat}{\dimexpr\sidemarg*\imgh/\striph\relax}          
    \setlength{\restA}{\dimexpr\imgw-\cutA\relax}
    \setlength{\restB}{\dimexpr\imgw-\cutA-\margnat\relax}
    \setlength{\startB}{\dimexpr\cutA-\margnat\relax}

\section{Elementary Quantum Operations and Pipelined Stabilizer Measurements}\label{app:gate_sequence}

    In this appendix, we describe the implementation of single and two-qubit gates, readout, and reset, as integrated into pipelined syndrome-extraction circuits \cite{Versluis2017} for the memory and teleportation experiments discussed in the main text.
    
    \begin{figure*}[t]
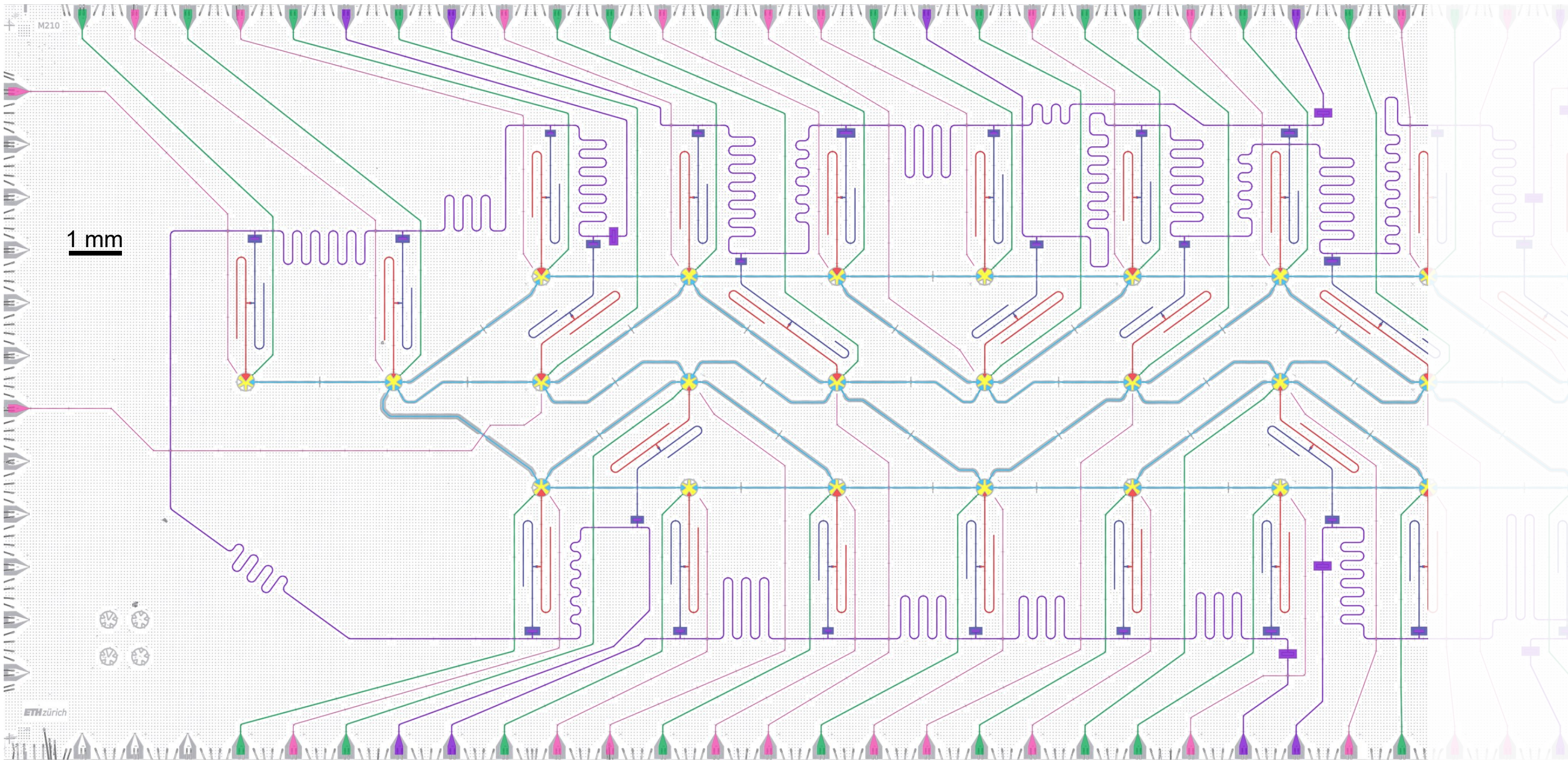

      \begin{minipage}[t][0.45\textheight][t]{\textwidth}
        \noindent\makebox[\textwidth][l]{%
          \includegraphics[trim={0 0 {\restA} 0}, clip, height=\striph]{figures/device.pdf}%
          {\transparent{0.35}%
           \includegraphics[trim={{\cutA} 0 {\restB} 0}, clip, height=\striph]{figures/device.pdf}}}%
        \par
        \caption{False-color optical micrograph of the left part of the quantum device. The grayed-out area is also displayed in \Cref{fig:s43_wide_2}. For legend, see \Cref{fig:fig1_quantum_device}(f).}
        \label{fig:s43_wide_1}
      \end{minipage}
    \end{figure*}

    We implement single-qubit $x$ and $y$ rotations using microwave pulses with a duration of \SI{32}{\nano\second}. 
    We use DRAG pulses~\cite{Motzoi2009, Gambetta2011a, Chen2016} with a cosine-squared envelope~\cite{Lazar2023a} to suppress spectral weight at the transition frequency between the first and second excited states, minimizing the leakage out of the computational subspace.
    The pulse amplitudes are calibrated using an error amplification method similar to the one discussed in Ref.~\onlinecite{Lazar2023}.
    To suppress gate errors and leakage from microwave crosstalk~\cite{Song2026e}, we apply cancellation pulses as in Refs.~\onlinecite{Krinner2022, Nuerbolati2022}.
    Single-qubit $z$ rotations are implemented as virtual $Z$ gates~\cite{McKay2017}.
    
    We realize C$Z$ gates using net-zero flux pulses that bring the $|11\rangle$ and $|20\rangle$ states of neighboring qubits into resonance~\cite{Rol2019, Negirneac2021}.
    During each gate, we apply flux pulses to both involved qubits, as in Ref.~\onlinecite{Krinner2022}.
    To yield a conditional phase of $\pi$, we adjust the idle time between the two rectangular halves of the net-zero pulse~\cite{Scarato2025}.
    We mitigate distortions of flux control pulses using infinite impulse response (IIR) and finite impulse response (FIR) filters, which we calibrate using the methods described in Refs.~\onlinecite{Rol2020, Hellings2025}.
    We compensate for flux crosstalk following Ref.~\onlinecite{Krinner2022}.
    The two-qubit gate durations, including buffer times, range from \SI{128}{ns} to \SI{160}{ns}.
    
    \begin{figure*}[t]
      \begin{minipage}[t][0.45\textheight][t]{\textwidth}
        \noindent\makebox[\textwidth][l]{%
          \hspace*{-\sidemarg}%
          {\transparent{0.35}%
           \includegraphics[trim={{\startB} 0 {\restA} 0}, clip, height=\striph]{figures/device.pdf}}%
          \includegraphics[trim={{\cutA} 0 0 0}, clip, height=\striph]{figures/device.pdf}}%
        \par
        \caption{False-color optical micrograph of the right part of the quantum device. The grayed-out area is also displayed in \Cref{fig:s43_wide_1}. For legend, see \Cref{fig:fig1_quantum_device}(f).}
        \label{fig:s43_wide_2}
      \end{minipage}
    \end{figure*}

    We perform three-level readout using frequency-multiplexed readout pulses~\cite{Blais2004, Chen2012f, Heinsoo2018}.
    During readout, the auxiliary qubits remain at their idle frequencies, and the data qubits are flux-tuned closer to their readout resonator frequencies to increase the dispersive shifts~\cite{Swiadek2024}.
    The readout pulse durations range from \SI{120}{ns} to \SI{240}{ns} for the auxiliary qubits and from \SI{240}{ns} to \SI{480}{ns} for the data qubits.
    Unconditional reset of the auxiliary qubits is implemented using 48-ns-long parametric flux pulses that swap the excited-state population of the qubit into the readout resonator mode, from where it decays into the feed line~\cite{Zhou2021, Lacroix2025}.
    
\begin{figure*}[htbp]
    		\centering
    		\includegraphics[width=\linewidth]{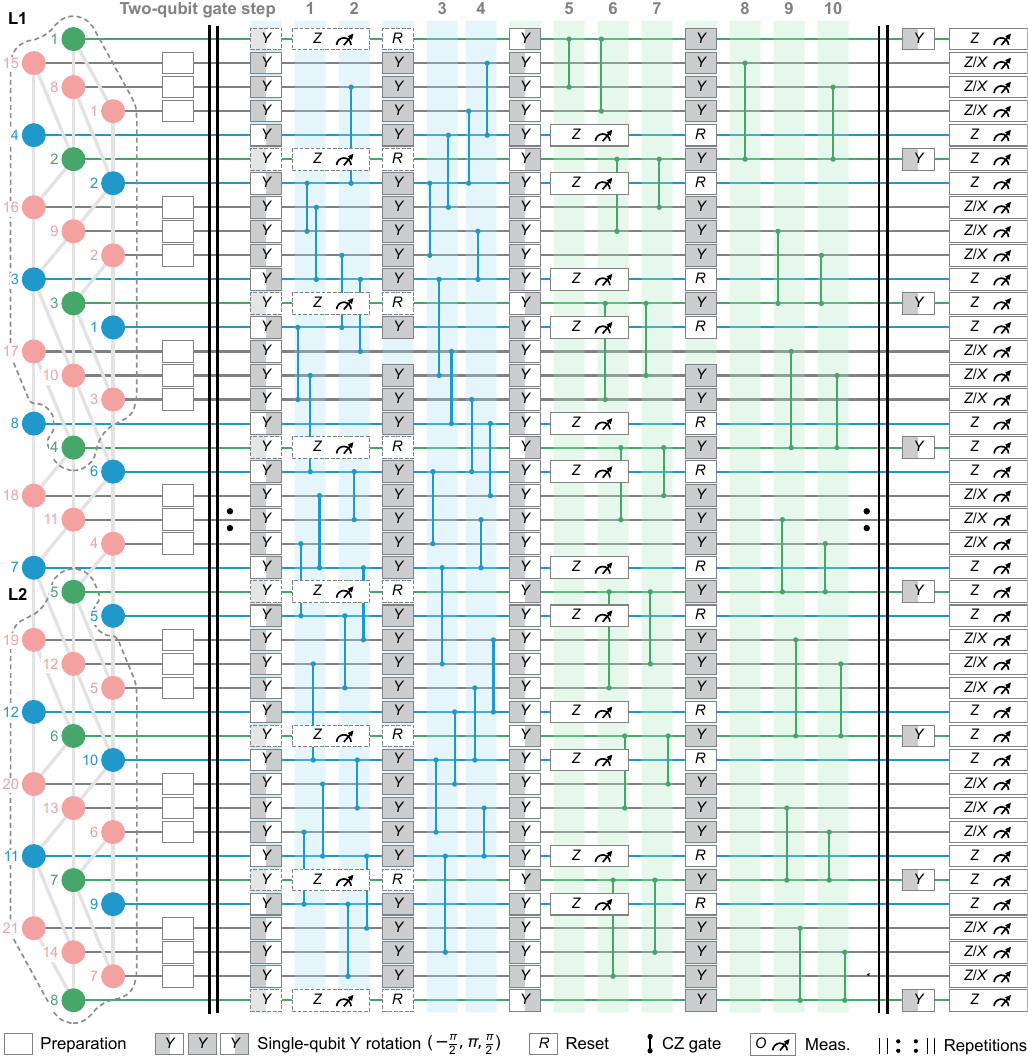}
    		\caption{\label{fig:gate_sequence}
    			Gate sequence of the memory experiment of the three-by-seven surface code.
                The qubits are shown on a deformed lattice, with gray lines indicating nearest-neighbor couplings.
                Red circles represent data qubits, while green (blue) circles represent $Z$-type ($X$-type) auxiliary qubits.
                Dashed outlines highlight the sublattices that correspond to the source (L1) and target (L2) qubits of logical teleportation experiments.
                Empty squares indicate single-qubit rotations on the data qubits for logical state initialization.
                The parallel execution of the $Z$-type ($X$-type) stabilizer circuits is indicated with green (blue) shaded backgrounds.
                Two-qubit gates are executed in ten time steps, labeled at the top of the diagram.
                The basis rotation, readout, and reset operations enclosed in dashed frames are omitted during the first cycle.
    		} 
    \end{figure*}

    We combine these elementary operations into syndrome-extraction circuits that form the basis of both the memory and teleportation experiments.
    As a representative example, we discuss the memory experiment of the three-by-seven surface code.
    The gate sequences are similar to the ones in Ref.~\onlinecite{Krinner2022}, with an additional reset of each auxiliary qubit after its mid-circuit measurement, see Fig.~\ref{fig:gate_sequence}.
    
    Before each experimental run, we read out all qubits (not shown in the figure) to determine their initial states, which we use for Pauli frame corrections on the final measurement outcome. Compared to ground-state heralding \cite{Riste2012, Johnson2012}, this procedure retains roughly twice as many experimental runs. However, it slightly increases error probabilities during the first rounds of stabilizer measurements due to the asymmetric error rate of the first excited state relative to the ground state, driven by $T_1$ decay.
    Then, we prepare the 21 data qubits in either of the two product states $|0\rangle ^{\otimes 21}$ and $|1\rangle ^{\otimes 21}$ ($|+\rangle ^{\otimes 21}$ and $|-\rangle ^{\otimes 21}$) to begin the process of initializing the cardinal logical qubit state $|0\rangle_\mathrm{L}$ or $|1\rangle_\mathrm{L}$ ($|+\rangle_\mathrm{L}$ or $|-\rangle_\mathrm{L}$).
    Each of the prepared product states is an equal superposition of equivalent instances of the target logical state.
    A single quantum error correction cycle then projects the data qubits into the eigenspace determined by the stabilizer outcomes measured in this cycle, deterministically initializing a logical state.
    
    In the memory experiment of the three-by-seven surface code, twelve $X$-type stabilizers (blue) and eight $Z$-type stabilizers (green) are active, see \Cref{fig:fig1_quantum_device}(c).
    We use a pipelined gate schedule~\cite{Versluis2017}, in which the extraction circuit for one stabilizer type begins concurrently with the measurement of the auxiliary qubits of the other type.
    Within each cycle, all active stabilizers of a given type are measured in parallel.
    In a pipelined scheme, all stabilizers can be measured in as few as eight steps, with four steps per stabilizer type~\cite{Versluis2017, Krinner2022}.
    We use four steps for the $X$-type stabilizers and six for the $Z$-type stabilizers.
    The two additional steps for gates of the $Z$-type stabilizers give us the freedom to choose C$Z$ gate interaction frequencies that avoid crossings with strongly coupled two-level-system defects~\cite{ColaoZanuz2025}.
    After the second time step of two-qubit gates, during $X$-type stabilizer extraction, we apply echo pulses to the data qubits and $X$-type auxiliary qubits to suppress dephasing from low-frequency flux noise, see Fig.~\ref{fig:gate_sequence}.
    We reset the $Z$-type auxiliary qubits to the ground state at the same time as these echo pulses.
    Similarly, after the seventh time step of two-qubit gates, during $Z$-type stabilizer extraction, we apply echo pulses to data qubits and $Z$-type auxiliary qubits, and we reset the $X$-type auxiliary qubits to the ground state.
    A complete QEC cycle, comprising $X$- and $Z$-type stabilizer measurements, has a duration of $1664\,\mathrm{ns}$.
    
    In the \teleportation{} protocols, we adapt the parallelization and timing of syndrome extraction for best success probabilities.
    First, we omit gates associated with inactive stabilizers at each step of the protocol.
    Second, to minimize residual-$ZZ$ errors, we halt stabilizer measurements during the split operation, in which flux pulses are applied to the data qubits during readout to increase readout fidelity.
    Finally, in the optimized teleportation protocol shown in Fig.~\ref{fig:fig4}(c), which measures only $X$-type stabilizers, we shorten the interval between consecutive $X$-type stabilizer measurements from $1008\,\mathrm{ns}$ to $256\,\mathrm{ns}$, limited by the readout holdoff time of the acquisition electronics.
    
\section{Leakage Rejection and Mean Syndrome Elements}\label{app:leakage}

\begin{figure}[!tb]
        \centering
        \includegraphics[width=\linewidth]{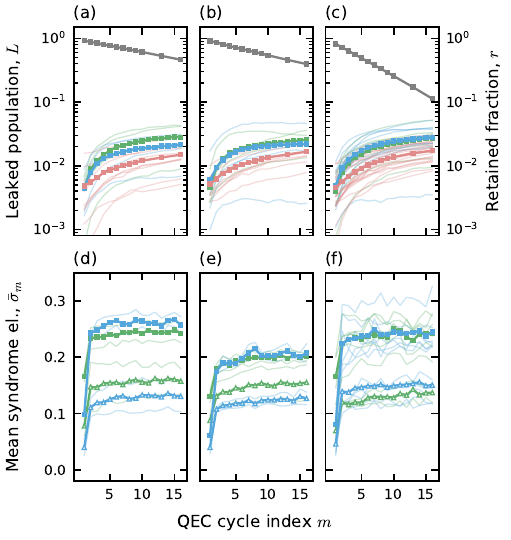}
        \caption{\label{fig:synd_and_leakage_rates}
            Leakage population and mean syndrome elements as a function of QEC cycle index $m$ on \Lone{} (a,d), \Ltwo{} (b,e), and \Lfull{} (c,f). 
            (a)--(c) Leakage populations on individual $X$-type (blue) and $Z$-type (green) auxiliary qubits and data qubits (red), shown as partially transparent lines, and average per qubit type, shown as opaque lines with squares.
            The overall retained fraction of runs after leakage rejection is shown in gray. 
            (d)--(f) Mean syndrome elements of individual $Z$-type (green) and $X$-type (blue) stabilizers (partially transparent lines), and average over weight-two (triangles) and weight-four (squares) stabilizers (opaque lines).
            }
    \end{figure}

    To investigate leakage rates and the evolution of mean syndrome elements across quantum error correction cycles, we analyze the memory experiments described in the main text. 
    For all experiments, we reject experimental runs where at least one qubit is measured in a leaked state~\cite{Krinner2022}. 
    This includes the measurement of all qubits at the beginning of the experiment, the mid-circuit readout of auxiliary qubits, and the final readout of all qubits. 
    We quantify the leakage population on auxiliary qubits from the mid-cycle measurements of a 16-cycle memory experiment.
    Since mid-circuit leakage of data qubits can only be detected in the final readout, we perform memory experiments with varying numbers of cycles to determine the leakage population of data qubits after $m$ cycles.
    We then average the measured leakage population over experiments with initializations in four cardinal states, $|0\rangle$, $|1\rangle$, $|+\rangle$, and $|-\rangle$.
    The leakage populations of individual qubits are shown as partially transparent curves for $Z$-type (green), $X$-type (blue), or data (red) qubits in \Cref{fig:synd_and_leakage_rates}(a)--(c). 
    This data is then averaged over all qubits of the same type, see opaque curves. 
    The per-cycle total probability of detecting leakage on any of the auxiliary qubits is \LeftAux{} and \RightAux{} for the individual logical qubits, and \largeAux{} for the $3 \times 7$ surface code. By comparison, the corresponding total probabilities for data qubits are \LeftDat{}, \RightDat{}, and \largeDat{}.
    We attribute the higher leakage rates of auxiliary qubits compared to data qubits to the implementation of two-qubit gates, which temporarily populates the $|2\rangle$-state of the auxiliary qubit~\cite{Negirneac2021}, see \Cref{app:gate_sequence}. 
    
    We fit an exponential model to the fraction of retained runs as a function of the number of cycles (gray squares) and obtain per-cycle retention fractions of ${r = 0.955}$, $0.945$, and $0.879$ for \Lone{}, \Ltwo{}, and \Lfull{}, respectively. 
    In the teleportation experiments in \Cref{sec:optimization}, the retained fractions averaged over the four cardinal initializations for the circuits in \Cref{fig:fig4}(a), (b) and (c) are $0.484$, $0.544$, and $0.719$, respectively.
    
    Syndrome elements $\sigma_m$ are obtained by comparing stabilizer values $s_m$ from consecutive rounds, ${\sigma_m = (1 - s_ms_{m-1})/2}$~\cite{Dennis2002, Fowler2012}. 
    A value of ${\sigma_m = 1}$ indicates a change in the stabilizer value, signaling the occurrence of an error. 
    For the memory experiments described in the main text, we calculate the $Z$-type ($X$-type) syndrome elements from experimental runs with initializations in the logical $Z$ ($X$) basis.  
    In all data sets, the mean syndrome element in the first cycle, $\bar\sigma_1$, is lower than the ones in subsequent cycles because it corresponds to a single stabilizer extraction measurement instead of to the comparison between two stabilizer extraction measurements. 
    Averaged over cycles $m \ge 2$ and across all stabilizers of a given weight, the mean syndrome elements for weight-two (weight-four) stabilizers are $0.14$ ($0.25$) for \Lone{}, $0.13$ ($0.20$) for \Ltwo{}, and $0.14$ ($0.24$) for \Lfull{}, see \Cref{fig:synd_and_leakage_rates}(d,e,f). These error probabilities exceed the lowest values reported in the literature~\cite{Acharya2025} by a factor of roughly $2.5$.
    
\section{Stabilizer simulations}\label{app:appendix_sims}

    We numerically simulate the memory and teleportation experiments of Figs.~\ref{fig:fig2_mean_syndromes}, \ref{fig:fig3_logical_state_teleportation}, and \ref{fig:fig4} using a stochastic Pauli noise model. Since the circuits consist only of Clifford gates, the protocols are efficiently simulable under Pauli noise~\cite{Aaronson2004}. All simulations are performed using Stim~\cite{Gidney2021}.
    
    Single- and two-qubit gate errors are represented by depolarizing channels with error probability $p$, applied immediately after the respective ideal gates:
    \begin{align*}
        \varepsilon^p_{1q}(\rho)
        &= (1-p)\rho
        + \frac{p}{3}\sum_{P \in \{X,Y,Z\}} P\rho P,
        \nonumber \\
        \varepsilon^p_{2q}(\rho)
        &= (1-p)\rho
        + \frac{p}{15}
        \sum_{\substack{
             P \in \{I,X,Y,Z\}^{\otimes 2}  \\
             P \neq I\otimes I 
        }}
        P\rho P.
    \end{align*}
    The error probabilities $p$ of the individual gates are inferred from the RB measurements summarized in \Cref{fig:fig2_mean_syndromes}(a).
    
    Readout errors are modeled as bit flips of the measurement outcomes. For each qubit, the bit flip probability is set to the mean state assignment error extracted from readout characterization measurements. Reset errors are modeled as an $X$ error applied after an ideal reset operation. For each qubit, the probability of such an error is given by the measured residual population averaged over qubits prepared in the ground or first excited state.
    
    To account for noise during idle intervals, we use the measured relaxation times $T_1$ and spin-echo coherence times $T_2^{\text{e}}$ of the individual qubits. For an idle interval of duration $t$, we apply the Pauli-twirled version of a combined amplitude- and phase-damping channel~\cite{Tomita2014a}, with Pauli error probabilities
    \begin{align*}
        p_x(t) &= p_y(t)
        = \dfrac{1}{4}\left(1-e^{-t/T_1}\right),
        \nonumber \\
        p_z(t) &=
        \dfrac{1}{2}\left(1-e^{-t/T_2^{\text{e}}}\right)
        - \dfrac{1}{4}\left(1-e^{-t/T_1}\right).
    \end{align*}

\section{Modular fault tolerance}\label{app:modular_FT}

    We call a distance-three module fault-tolerant if (i) in the absence of internal faults, i.e., faults within the module, it corrects every single-qubit physical error on the input state and implements the intended logical operation; and (ii) for an error-free input, any single internal fault preserves the intended logical operation up to a correctable residual error on the output state. A fault-tolerant module must satisfy these criteria even when decoding using only syndrome information generated within the module. In the teleportation experiments, we nevertheless decode the syndrome data globally, as matching-based decoding at the module level introduces additional challenges~\cite{Skoric2023,Vuillot2019}. We leave the investigation of modular decoding schemes to future work. For a more general definition of fault tolerance, we refer to~\cite{Gottesman2024}, noting that this is overly restrictive for the surface code syndrome extraction circuit~\cite{Old2025a}.

    We first discuss the fault tolerance of the fully modular circuit, see \Cref{fig:fig3_logical_state_teleportation}(c) and the more detailed schematic in \Cref{fig:fig4}(a), before discussing the fault tolerance of the optimized circuits in \Cref{fig:fig4}(b) and (c).
    The initialization module has no input; therefore, all faults are treated as internal faults. Any single fault during physical qubit preparation---i.e., initial readout and single-qubit rotations---or during stabilizer measurement leaves the logical qubit in the intended logical state up to a correctable residual error. Although a single round of stabilizer measurements would suffice to satisfy this fault-tolerance criterion for a distance-three code, we perform two rounds, the second of which provides temporal syndrome comparisons for error correction within the initialization module. Interchanging the $\hat{S}^X$ and $\hat{S}^Z$ stabilizer measurement layers does not affect fault tolerance. In the protocols of Fig.~\ref{fig:fig4}(a) and (b), we begin with $\hat{S}^X$ stabilizer measurements, whereas in the protocol of Fig.~\ref{fig:fig4}(c), we begin with $\hat{S}^Z$ stabilizer measurements to enable parallelization of subsequent rounds of stabilizer measurements.
    
    In the merge-and-split module, we perform three rounds of stabilizer measurements. The $\hat{S}^X$ stabilizer measurements protect the inferred joint logical parity against a single stabilizer measurement error and provide a timelike distance of three. The $\hat{S}^Z$ stabilizer measurements detect single-qubit physical errors in the $Z$ basis that affect the input state. Two rounds of $\hat{S}^Z$ stabilizer measurements would be sufficient to distinguish physical errors on the input state from stabilizer measurement errors. A third round is included to maintain the alternating sequence of $\hat{S}^Z$ and $\hat{S}^X$ stabilizer measurement rounds. The additional rounds of stabilizer measurements on \Lone{} and \Ltwo{} during the split are not required for fault tolerance either, but we include them to demonstrate mid-circuit data qubit readout while preserving the individual logical qubits.
    
    In the measurement modules, logical $X$- and $Z$-basis measurements are implemented transversally and are fault-tolerant because the data qubit measurement outcomes provide both the logical measurement result and the stabilizer values in the respective basis. A single readout error can therefore be treated as an effective data qubit error and can be corrected when reconstructing the logical outcome.
    
    Through circuit optimization, the target-qubit initialization, merge-and-split, and source-qubit measurement modules are combined into a single fault-tolerant expand-and-shrink module, see \Cref{fig:fig4}(b). In this composite module, the three stabilizer measurement rounds provide enough syndrome information about errors on the target patch. Separate stabilizer measurement rounds are therefore not required for the target-qubit initialization. The stabilizer measurement rounds on the source patch during the split in \Cref{fig:fig4}(a) are also unnecessary to ensure fault tolerance, as stated above; we therefore omit them in the optimized circuit in \Cref{fig:fig4}(b), noting that the source qubit
    does not need to persist after the split operation. 
    
    In a second optimization step, we further reduce the number of stabilizer measurements by depleting~\cite{Bodeker2026} the expand-and-shrink module, see \Cref{fig:fig4}(c). We retain the $\hat{S}^X$ stabilizer measurements needed to reconstruct the outcome of the \XSXT{} measurement fault-tolerantly, but we omit the $\hat{S}^Z$ stabilizer measurements as internal faults in the $Z$ basis do not have to be corrected within the module. The $Z$-basis readout of the source patch supplies both the logical measurement outcome and the $Z$-type stabilizer values. These values allow single input errors in the $Z$ basis on the source patch to be corrected when reconstructing the logical measurement outcome, without additional $\hat{S}^Z$ stabilizer measurements within the expand-and-shrink module. The target qubit is initialized within the module and therefore has no input errors. Internal faults in the $Z$ basis can only propagate to correctable residual errors on the target qubit, as these cannot alter the measurement outcome of \XSXT. Consequently, $\hat{S}^Z$ measurement rounds can also be omitted for the target logical qubit \Ltwo{}. Moreover, we measure the previously established $X$-type stabilizers on the source patch only twice within the module. The expected values of these stabilizers are already known at the input of the module, as the module has access to the updated Pauli frame. 
    Because updating the Pauli frame is equivalent to virtually applying a correction after the preceding module, using this information for decoding does not constitute out-of-module information.
    The outcomes of the first measurement round of $\hat{S}^X$ stabilizers is therefore compared with this reference to detect errors. In contrast, stabilizers on the other patches lack this reference because their initial values are random, requiring three rounds of $\hat{S}^X$ stabilizer measurements.

\section{Time evolution of logical observables when teleporting $\ket{1}_\Lone{}$ and $\ket{-}_\Lone{}$}\label{app:time_evolution_appendix}

\begin{figure}
        \centering
        \includegraphics[width=\linewidth]{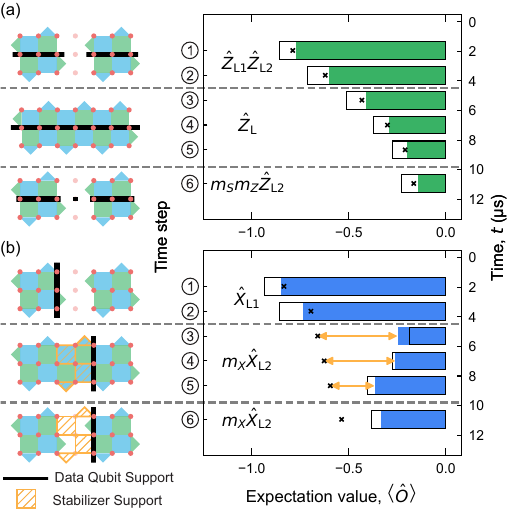}
        \caption{Time evolution of logical observables during \teleportation. Pictograms of the surface code lattice showing the active stabilizers at the steps in the circuit in \Cref{fig:fig3_logical_state_teleportation}(c).
        Support of the logical observable is indicated by black lines and orange patches, see legend. Colored bars indicate error-corrected expectation values of the corresponding observable at each step. Wireframes show results of simulations. Black crosses indicate calculations based on errors only on the data qubit support, see \Cref{app:spacelike_model}. (a) Teleportation of $\ket{1}_\Lone{}$. (b) Teleportation of $\ket{-}_\Lone{}$. Orange arrows are differences between these calculations and the measured data.}
        \label{fig:time_evolution_appendix}
    \end{figure}

    In addition to evaluating the time evolution of the logical $Z$ and $X$ observables using $\ket{0}_\Lone{}$ and $\ket{+}_\Lone{}$ in the circuit in \Cref{fig:fig3_logical_state_teleportation}(c), we also measure their evolution when teleporting the $-1$ eigenstates $\ket{1}_\Lone{}$ and $\ket{-}_\Lone{}$.
    The observables carrying the logical information are the same as for the $+1$ eigenstates and are replicated in the pictograms in \Cref{fig:time_evolution_appendix}(a,b).
    As expected for ideal target values of $-1$ rather than $+1$, the final expectation values have comparable magnitudes but opposite signs, with $\langle m_sm_Z\ZT\rangle=\ZstepSixAppendix$ and $\langle m_X\XT{}\rangle=\XstepSixAppendix$, see the green and blue bars for the last time step in \Cref{fig:time_evolution_appendix}(a,b). These values correspond to success probabilities of \ZsuccessApp{} for $\ket{1}_\Lone{}$ and \XsuccessApp{} for $\ket{-}_\Lone{}$.
    Moreover, the time evolutions closely match those observed for the $+1$ eigenstates. The magnitude of the expectation value of the $Z$ observable decreases steadily, while the evolution of the expectation value of the $X$ observable exhibits the same distinct pattern observed in \Cref{fig:fig3_logical_state_teleportation}(e).
    
\section{Expectation values calculated from errors on the data qubit support}\label{app:spacelike_model}    
    In this appendix, we calculate the logical $X$ and $Z$ expectation values for each time step in the circuit in \Cref{fig:fig3_logical_state_teleportation}(c) when only considering contributions from errors on the data qubit support.
    The expectation value for a given time step~\Circled{$i$} can be calculated from the one in the preceding time step and the logical error probability $\epsilon_i$ in step~\Circled{$i$}. 
    With probability $1 - \epsilon_i$, no logical error occurs, leaving the sign of the observable as in the previous step. Thus, the expectation value at step $i$ is given by
    \begin{align*}
        \langle O \rangle_i &= (1 - \epsilon_i) \langle O \rangle_{i-1} + \epsilon_i \left(-\langle O \rangle_{i-1}\right) \\&= (1 - 2\epsilon_i) \langle O \rangle_{i-1}\\
        &=\prod_{j=1}^{i}(1 - 2\epsilon_j) \langle O \rangle_{0}.
    \end{align*}
    The logical error probability $\epsilon_i$ is determined by the logical observable that carries the logical information in step~\Circled{$i$}, see the black lines and orange patches in the pictograms of \Cref{fig:fig3_logical_state_teleportation}(d) and (e). 
    We express each ${\epsilon_j=\epsilon_j^\mathrm{D}\left(1-\epsilon_j^\mathrm{S}\right)+\epsilon_j^\mathrm{S}\left(1-\epsilon_j^\mathrm{D} \right)}$ by separating errors on the data qubit support $\epsilon_j^\mathrm{D}$ from those on the stabilizer measurement support $\epsilon_j^\mathrm{S}$. To compute the expectation values shown as black crosses in Figs.~\ref{fig:fig3_logical_state_teleportation}(d,e) and \ref{fig:time_evolution_appendix}(a,b), we consider only errors on the data qubit support, i.e., $\epsilon_j^\mathrm{S}=0$ and thus $\epsilon_j=\epsilon_j^\mathrm{D}$, which we obtain $\epsilon_j^\mathrm{D}$ from the memory experiments in \Cref{fig:fig2_mean_syndromes}(b) and (c). While the error probabilities are determined by the three-by-seven surface code for steps~\Circled{3} to \Circled{5}, they are given by combinations of the logical error probabilities of the individual logical qubits \Lone{} and \Ltwo{} for steps~\Circled{1}, \Circled{2}, and \Circled{6} as shown in \Cref{tab:space-like}. 

    \begin{table}[t]
        \centering
        \setcellgapes{5pt}
        \setlength{\tabcolsep}{0.33em}
        \begin{tabular}{ccccc}
        Step & $Z$ & $\epsilon_Z$& $X$& $\epsilon_X$ \\ 
        \toprule 
        \Circled{1}--\Circled{2} & \ZSZT{} & $\begin{aligned}[t]
          & \hphantom{{}+{}} \epsilon_{Z_\Lone}(1-\epsilon_{Z_\Ltwo}) \\
          & {} + \epsilon_{Z_\Ltwo}(1-\epsilon_{Z_\Lone}) \\
          & \approx \leftZRightZ
        \end{aligned}$ & \XS & $\epsilon_{X_\Lone} \approx \leftX$ \\ 
        \midrule 
        \Circled{3}--\Circled{5} & $\hat{Z}_\Lfull{}$& $\epsilon_{Z_\Lfull{}} \approx \fullZ$ & $m_X$\XT{} & $\epsilon_{X_\Lfull{}} \approx \fullX$ \\ 
        \midrule 
        \Circled{6} & $m_Sm_Z$\ZT{} & $\begin{aligned}[t]
          & \hphantom{{}+{}} \epsilon_{Z_\Lone}(1-\epsilon_{Z_\Ltwo}) \\
          & {} + \epsilon_{Z_\Ltwo}(1-\epsilon_{Z_\Lone}) \\
          & \approx \leftZRightZ
        \end{aligned}$ & $m_X$\XT{} & $\epsilon_{X_\Ltwo} \approx \rightX$ \\ 
        \end{tabular}
        \caption{Logical $Z$ and $X$ observables at each step of the teleportation circuit in \Cref{fig:fig3_logical_state_teleportation}(c), along with the logical error probabilities $\epsilon_Z$ and $\epsilon_X$ arising from errors on the data-qubit support of each observable.}\label{tab:space-like}
    \end{table}
    
\section{Basis-dependent performance gains from circuit optimization}\label{app:basis_dependence}
    
    The first optimization step discussed in \Cref{sec:optimization} improves the logical fidelity in the $Z$ basis, but not in the $X$ basis. 
    This improvement is achieved by removing stabilizer measurements of the target qubit after initialization and of the source qubit before measurement, compare the circuits in \Cref{fig:fig4}(a,b).
    \begin{figure}[b]
    	\centering
    	\includegraphics[width=\linewidth]{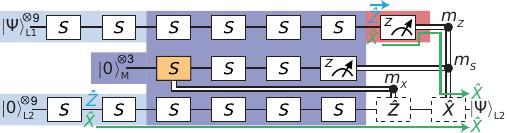}
    	\caption{Propagation (colored arrows) of logical $\hat{X}$ (green) and $\hat{Z}$ (blue) errors occurring in the source-qubit measurement and target-qubit initialization modules. Empty white and orange boxes represent stabilizer measurements. See legend in \Cref{fig:fig3_logical_state_teleportation}(c) for other elements.}
    	\label{fig:error_propagation}
    \end{figure}
    Logical errors in the $Z$ basis of the target qubit, i.e., logical $\hat{X}_\Ltwo{}$ errors, before the merge operation commute with all stabilizers and thus remain logical $\hat{X}_\Ltwo{}$ errors until the end of the protocol. A logical $\hat{X}_\Lone{}$ error on the source qubit after the split changes the conditional Pauli operation $\hat{X}_\Ltwo$, see green arrows in \Cref{fig:error_propagation}. Consequently, removing these segments of the circuits is expected to improve the fidelity in the $Z$ basis. 
    In the $X$ basis, logical $\hat{Z}_\Ltwo{}$ errors on the target qubit before the merging step have no effect on the initial state $\ket{0}_\Ltwo{}$.
    Finally, logical $\hat{Z}_\Lone{}$ errors on the source qubit after the split do not affect the $Z$-basis measurement and the Pauli frame update conditioned on it, see the blue arrow in \Cref{fig:error_propagation}. Thus, the fidelity in the $X$ basis remains unaffected by the omission of these circuit segments.

\end{document}

%% file: memory_variables.tex
\newcommand{\fullZ}{\ensuremath{\SI{15.2}{\percent}}}
\newcommand{\leftZ}{\ensuremath{\SI{6.3}{\percent}}}
\newcommand{\rightZ}{\ensuremath{\SI{4.9}{\percent}}}
\newcommand{\fullX}{\ensuremath{\SI{2.5}{\percent}}}
\newcommand{\leftX}{\ensuremath{\SI{8.3}{\percent}}}
\newcommand{\rightX}{\ensuremath{\SI{4.9}{\percent}}}
\newcommand{\leftZRightZ}{\ensuremath{\SI{10.6}{\percent}}}

%% file: gate_variables.tex
\newcommand{\SQG}{\ensuremath{\num{1.2(9)e-3}}}
\newcommand{\TQG}{\ensuremath{\num{1.6(8)e-2}}}
\newcommand{\ROtwo}{\ensuremath{\num{1.3(12)e-2}}}
\newcommand{\ROthree}{\ensuremath{\num{2.2(10)e-2}}}
\newcommand{\reseterror}{\ensuremath{\num{6(5)e-3}}}
\newcommand{\weightTwo}{\ensuremath{\num{4.4(17)e-2}}}
\newcommand{\weightFour}{\ensuremath{\num{7.5(13)e-2}}}

%% file: protocol_variables.tex
\newcommand{\fidTeleportX}{\ensuremath{0.32}}
\newcommand{\fidTeleportZ}{\ensuremath{0.14}}
\newcommand{\fidMoveX}{\ensuremath{0.30}}
\newcommand{\fidMoveZ}{\ensuremath{0.23}}

\newcommand{\fidDepletedX}{\ensuremath{0.47}}
\newcommand{\fidDepletedZ}{\ensuremath{0.29}}

\newcommand{\successDepletedX}{\ensuremath{0.73}}
\newcommand{\successDepletedZ}{\ensuremath{0.65}}
\newcommand{\ModularToMoveZ}{\ensuremath{0.09}}

%% file: step_variables.tex
\newcommand{\ZstepOne}{\ensuremath{0.77}}
\newcommand{\ZstepSix}{\ensuremath{0.14}}
\newcommand{\XstepOne}{\ensuremath{0.84}}
\newcommand{\XstepTwo}{\ensuremath{0.72}}
\newcommand{\XstepThree}{\ensuremath{0.23}}

\newcommand{\XstepFive}{\ensuremath{0.36}}
\newcommand{\XstepSix}{\ensuremath{0.31}}
\newcommand{\ZstepSixAppendix}{\ensuremath{-0.15}}
\newcommand{\XstepSixAppendix}{\ensuremath{-0.33}}
\newcommand{\Zsuccess}{\ensuremath{0.57}}
\newcommand{\Xsuccess}{\ensuremath{0.65}}
\newcommand{\ZsuccessApp}{\ensuremath{0.57}}
\newcommand{\XsuccessApp}{\ensuremath{0.67}}

%% file: leakage_rejection.tex
\newcommand{\largeAux}{\ensuremath{\num{1.105(11)e-1}}}
\newcommand{\largeDat}{\ensuremath{\num{1.99(10)e-2}}}
\newcommand{\LeftAux}{\ensuremath{\num{4.15(3)e-2}}}
\newcommand{\LeftDat}{\ensuremath{\num{6.6(4)e-3}}}
\newcommand{\RightAux}{\ensuremath{\num{4.972(18)e-2}}}
\newcommand{\RightDat}{\ensuremath{\num{7.6(4)e-3}}}